# Cuban natural palygorskite nanoclays for the removal of sulfamethoxazole from aqueous solutions

D. Hernández[1†], L. Quiñones[1†], L. Lazo[1], C. Charnay[2], M. Velázquez[3], A. Rivera[1*]

[1]*Zeolites Engineering Laboratory, Institute of Materials Science and Technology (IMRE), University of Havana, Cuba*

[2]*Institut Charles Gerhardt de Montpellier, CNRS UMR 5253, Université de Montpellier, CNRS, ENSCM, Place Eugène Bataillon, 34095, Montpellier cedex 05, France*

[3] *Research Center for Mining and Metallurgical Industries (CIPIM), La Havana, Cuba*

† The authors contributed equally to this work
*Corresponding author: aramis@imre.uh.cu

## Abstract

Water pollution with pharmaceutical and personal care products has become a serious environmental. A reasonable strategy to mitigate the problem involves absorbent materials. In particular, the use of natural clays is an advantageous alternative considering their high adsorption capacity and compatibility with the environment. In the present work, the efficacy of a Cuban natural clay (palygorskite, Pal) as support of sulfamethoxazole (SMX) ―an antibiotic considered an emerging contaminant (EC)― was evaluated. The amount of SMX incorporated onto clay was determined by UV spectroscopy. The resulting composite material was characterized by infrared spectroscopy (IR), X-ray diffraction (XRD), thermogravimetric analysis (TG/DTG), zeta potential (ZP), nitrogen adsorption measurements and transmission electron microscopy (TEM). The drug desorption studies in aqueous solution indicated the reversibility of the incorporation process, suggesting the potential use of the Pal nanoclay as an effective support of SMX and hence, a good prospect for water decontamination.

## Introduction

Pollution has important implications in the society due to its effects on the aquatic environment and human health. The so-called emerging contaminants deserve special attention, partly because many of them are chemical products of common use. The risk they pose to human health and the environment is not yet fully understood. In the case of antibiotics, high concentrations in wastewater may induce drug-resistance in animals and human beings that eventually consume that water. The overall result is bacterial immunity to treatments with commonly used antibiotics. The effectiveness of many wastewater treatment systems cannot be trusted, since they are not monitored due to the absence of specific regulations for these contaminants [1]. To face this problem, science and technology are opting for a set of physical, chemical and biological methods for environmental remediation [2-4], among which the adsorption stands out for their simplicity and easy application.

Within the list of adsorbent materials, clays play an important role for their demonstrated adsorptive properties [5-10]. The use of natural clays for environmental application has shown to be a viable and economical alternative [11,12]. Different families of clays like smectite, sepiolite, kaolinite and palygorskite [13-19] have been used as pollutant adsorbents, especially in the case of emerging contaminants. Natural Palygorskite is a fibrous clay with structure 2:1 where the tetrahedral silica sheets are periodically inverted with respect to the tetrahedral bases, leading to a periodically interruption in the octahedral sheets, and the cations occupying terminal positions must complete their coordination sphere with water molecules [5].

A typical emergent contaminant is the antibiotic sulfamethoxazole, SMX (4-amino-N-(5-methyl-3-isoxazolyl) benzenesulfonamide), which belongs to the sulfonamides family [20]. It is an antibiotic of wide spectrum used in humans, animals and agriculture, whose presence is common in the wastewater and soils [21]. This drug is prescribed to treat respiratory, urinary tract, skin and gastrointestinal infections. It is also widely used in aquaculture and livestock breeding for animal health and in some countries as a growth promoter [22]. SMX is listed within the top 30 most frequently detected wastewater contaminants, having a half-life of 85 to 100 days and more [22,23]. Moreover, this emerging contaminant has been regarded by many authors as a persistent pollutant [23], which can induce high levels of bacterial resistance [24].

Inspired by the principles of green chemistry, the general objective of the present work is proposing a new strategy for the remediation of water from emerging and persistent contaminants by using a Cuban natural nanoclay ─an abundant and inexpensive material─ combined with a simple methodology. Here, the capture of SMX by a Cuban natural palygorskite (Pal) was evaluated. The drug-clay composites were studied by different characterization techniques (XRD, IR, TG, ZP and nitrogen adsorption). Desorption studies in water were carried out in order to demonstrate the reversibility of the incorporation process. Our results strongly indicate that Cuban natural palygorskite is an excellent candidate for the removal of SMX from water.

## Experimental

**Materials**

Cuban palygorskite (Pal) was used as raw material, and it was supplied by the Research Center for the Mining-Metallurgical Industry (CIPIMM). It contains about 57% Pal, 32% montmorillonite (Mt), 11% of quartz (Q)[25]. Sulfamethoxazole (SMX) in the neutral molecular form ($C_{10}H_{11}N_3O_3S$) was the model drug employed like emerging contaminant. It was chosen considering its widespread consumption and occurrence in water bodies and soils. SMX was used as received from the Cuban pharmaceutical industry. The chemical structure of different species of the drug as a function of pH is show in figure 1.

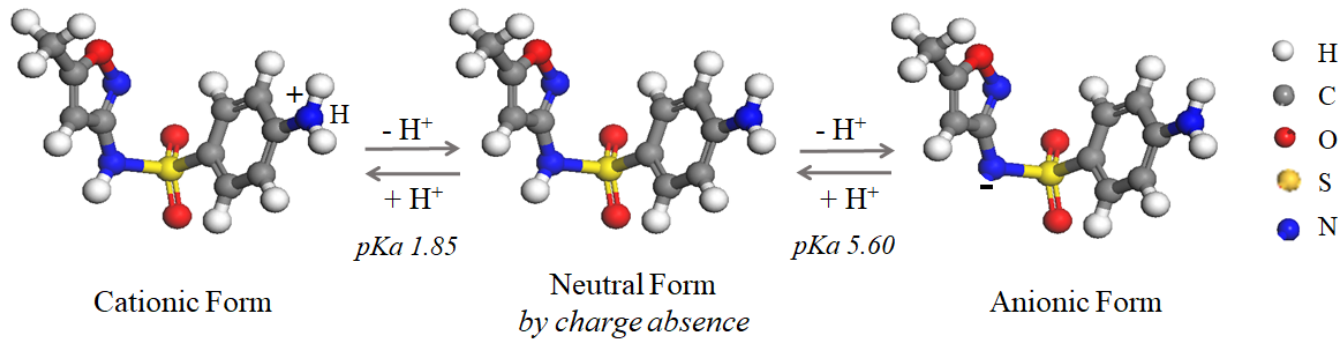

Figure 1 Acid-base equilibrium of the SMX molecule.

**Interaction procedure**

The study was performed using a liquid-solid ratio of 10:100 (3 mg/mL of drug aqueous solution per mg of Pal powder) under magnetic stirrer by 3 h at room temperature and acid pH. After the interaction, the dispersions were centrifuged during 15 min at 300 rpm. The resulting material was dried in an oven at 65°C during 24 h. The drug concentration in the supernatant before and after interaction were measured by UV spectroscopy at an adsorption maximum of 265 nm, using a spectrophotometer Rayleigh UV-2601equipped with a quartz cell having a path length of 0.1 cm and a resolution of 1 nm.

The drug uptake (i.e., load adsorbed by the clay) was as:

$$q_e = \frac{(C_o - C_f) \times V}{m} \quad (1)$$

where $q_e$ (mg/g) is the drug mass adsorbed per unit mass of the adsorbent, $C_o$ is the initial drug concentration in solution (mg/mL), $C_f$ is the final drug concentration in solution (mg/mL), $V$ is the solution volume (mL) and $m$ is the adsorbent mass (g) used in the experiments.

All the tests were performed for triplicate for each sample, and the average values were used in data analysis. The maximum difference between the outputs and their average was of 10 mg/g, which corresponds to a relative uncertainty of around of 3 % in the mass of adsorbed drug.

## Characterization

X-ray diffraction (XRD) patterns of the solid samples in powder form were collected by a Philips Xpert diffractometer with Cu K$\alpha$ radiation ($\lambda$ = 1.54 Å) at room temperature, operating at a voltage of 45 kV and a working current of 25 mA. The experiments were done at a scan rate of 1° min$^{-1}$ for a 2θ range spanned from 2 to 70°.

Fourier transform infrared spectra (FTIR) were measured using a Nicolet AVATAR 330 spectrometer in the wavenumber interval of 400-4000cm$^{-1}$, with a 2 cm$^{-1}$ resolution. The raw materials and the composites were prepared using KBr pressed-disk technique with 0.8% inclusion of the material to be analyzed.

Thermogravimetric (TGA) analysis was carried out within 25–800°C at the heating rate of 10°C/min under dry nitrogen flow (20 mL/min) using a Perkin Elmer STA6000. The sensitivity of the thermos balance was ±1 µg. The solid sample mass used in each test was about 30mg.

The surface charge of the particles was evaluated employing a zeta potential analyzer (Malvern Nano Zetasizer instrument). 1 mg of each sample, palygorskite or Pal-SMX composite, was dispersed fully in 2 mL of a KCl solution at 0.001 M using an ultrasonic bath for 15 min at room temperature. The pH was adjusted in the range 2.0–10.0 pH units employing NaOH or HCl solutions at 2 M.

Nitrogen adsorption-desorption experiments were performed at liquid nitrogen temperature (−196 °C) on a Micromeritics TriStar 3020 surface area and porosity analyzer. Samples were outgassed for 8 h under high vacuum at 100 and 150°C for composite and clay, respectively. Surface areas were determined using the Brunauer–Emmett–Teller (BET) method.

Transmission electronic microscopy (TEM) images were taken by a JEOL 1200EX2 equipped with a CDD captor of 11 Mpixels (SIS Olympus camera Quemesa model) and an acceleration tension of 100 kV. For the TEM samples preparation, a suspension of a small amount of sample in ethanol was given an ultrasonic treatment for 5 min and then dropped onto a grid.

**Drug desorption studies**

Desorption studies were carried out with the objective of evaluating the reversibility of the SMX incorporation process onto Pal clay. The method used is similar to that reported by Blasioli et al. [26]. Desorption was performed on loaded Pal obtained after the drug adsorption by substituting one SMX solution half-volume with distilled water. Then, the clay-drug suspension was stirred for 24 h, centrifuged, and the supernatant analyzed by UV. After that, the half volume of supernatant was again removed and substituted with fresh distilled water. This dilution step was repeated several times at intervals of 24 h during 96 h. The procedure was performed by triplicate. Desorption percent was calculated as follows:

$$Drug\ desorption\ \% = \frac{unloaded\ SMX\ (g)}{loaded SMX\ (g)} \times 100\%$$

**Results and discussion**

*Drug incorporation and samples characterization*

The experimental results evidenced the SMX incorporation onto Pal, $260 \pm 10$ mg/g for an efficiency of 88 %. This could be explained based on the acid and basic equilibrium of the drug: the incorporation seems to take place when the drug is in its neutral form. It suggested that the interaction between drug and clay occurs through weak interactions, i.e., Van der Waals forces and/or hydrophobic interactions [20].

XRD patterns of the Pal and the clay-drug composite (Pal-SMX) are shown in Fig.2. According to the most intense reflections, the analysis confirmed that the Pal is the main mineralogical phase in the samples [27-29]. Other phases associated were also identified as montmorillonite (Mt) and quartz (Q) [27]. In the Pal pattern, the maxima in the called "Chisholm zone" at $2\theta=19.8º$, $20.78º$ and $22.08º$, are associated with a crystalline system [5,30], corresponding to the 040, 121 and 310 Pal basal reflections, respectively. The last one coincides with the 100 reflection of the Q present in the sample. In natural clays, the presence of impurities like Q (maximum at $20.78º$, *d*-value calculated according to Bragg's Law for this angle was 0.43nm) does not allow to discriminate if the palygorskite is mainly orthorhombic, monoclinic or a mixture of both phases [31].

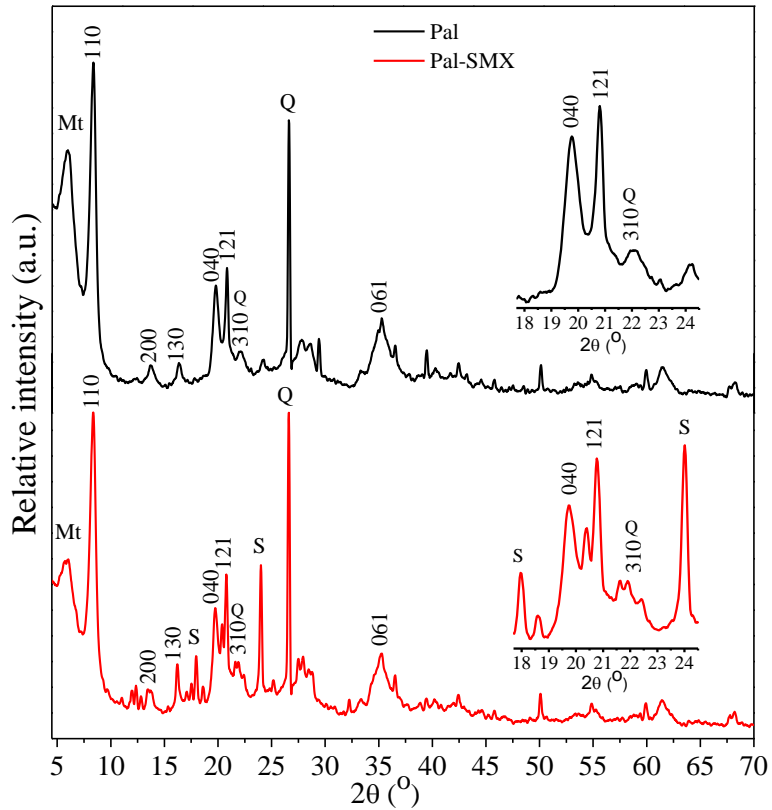

Figure 2. XRD patterns of the Pal and the Pal-SMX composite. The *hkl* values of the characteristic planes of each phase are indicated. The insets zoom the Chisholm zone.

In the Pal-SMX pattern, the characteristic reflections of the Pal phase were identified. New signals attributed to the presence of SMX (labeled as S) on the composite were observed (see Fig. 2) [32]. No significant variation in the interplanar distance values of the reflections associated to the Pal phase were detected, which can be explained in a simple way considering the clay structure and the dimensions of its channels, as well as the SMX molecule size. The Pal is a mineral of rigid structure with channels size around of 0.4 x 0.6 nm along $c$ axis [25], while the SMX is an organic molecule with bigger dimensions—0.89 nm×0.80 nm×0.72 nm [20]—than those of the Pal channels. Thus, the entrance and diffusion of this drug inside of Pal structure is not probable, which suggests that the SMX molecules are adsorbed onto the clay surface. The diffraction pattern indicates no structural changes in the material after the drug adsorption process, confirming its long range order stability. In addition, no important variations in the position of 001 basal reflection of Mt was detected. It indicates the no intercalation of SMX into Mt, which corroborates that drug molecules are adsorbed onto the material surface.

FTIR spectra of the starting materials (Pal, SMX) and the Pal-SMX composite are shown in Fig.3. In the Pal spectrum, different bands corresponding to stretching vibration O−H ($v_{OH}$) associated to several type of water present in the structure of the clay appear in wavenumber region of 3800 at 2800 cm$^{-1}$. The signal at 3614 cm$^{-1}$ was attributed to the stretching vibrations of the hydroxyl groups bonded to the cations in the lattice (Al$_2$−OH fundamentally and Al−Fe$^{3+}$−OH to a lesser extent), with octahedral coordination. This position and intensity of the structural water band is related with the content of R$^{3+}$ cations and the dioctahedral sites in the clay's structure [33-35]. The band centered at 3580 cm$^{-1}$ was assigned to coordinated water in the channels [34]. However, other authors attributed this band to Al−Fe$^{3+}$−OH or Al−Mg−OH bondsm[33,36]. The signals at 3550, 3406 and the shoulder at 3260 cm$^{-1}$ were assigned to zeolitic and coordinated water [35]. The band at 1651 cm$^{-1}$ corresponds to the O−H characteristic bending vibration ($\delta_{OH}$) of the zeolitic water, and/or the adsorbed water [35,36]. In the region between 1200 and 400 cm$^{-1}$ the characteristic signals of Si−O stretching modes of the tetrahedral sites and Al−OH or Mg−OH deformation can be detected (see Fig. 3) [36].

The bands observed at 1194, 1120 and 1030 cm$^{-1}$ correspond to the stretching vibrations Si−O−Si ($v^{as}_{Si-O-Si}$) [34,35]. There are three types of Si−O−Si links in Pal forming different connections with other atoms of the crystal structure: the first (labeled Si–O$_1$–Si) observed at 1194 cm$^{-1}$ connects the two inverse tetrahedrons (SiO$_4$). This signal is characteristic of palygorskite and sepiolite clays (whose structure are constituted by channels), and its presence is due to the inverted tetrahedrons through the apical oxygens. It does not appear in the IR spectrum of the clay. The second type (denoted Si−O$_2$−Si) appears at 1120 cm$^{-1}$ and it connects two SiO$_4$ tetrahedrons in the two pyroxene chains. The third connection type (labeled Si–O$_3$–Si) associated to the vibration at 1030 cm$^{-1}$ connects the two SiO$_4$ tetrahedrons in a simple chain of pyroxene [37,38]. The band at 986 cm$^{-1}$ is attributed to SiO$_4$ tetrahedron base breathing stretching vibration and the perpendicular Si−nonbridging oxygen−Mg (Si–O$_{nb}$–Mg) asymmetric stretching vibrations. At 876 and 506 cm$^{-1}$ it can be observed the Al−OH−Mg and Si−O−Al vibrations, respectively. This last band is related with the octahedral Al content [35]. The signal at 645 cm$^{-1}$ is characteristic of Si–O$_{nb}$–Si stretching ($\delta_{Si-O_{nb}-Si}$) and the band at 481 and 470 cm$^{-1}$ it is attributed to bending vibration O−Si−O ($\delta_{O-Si-O}$) [37,39].

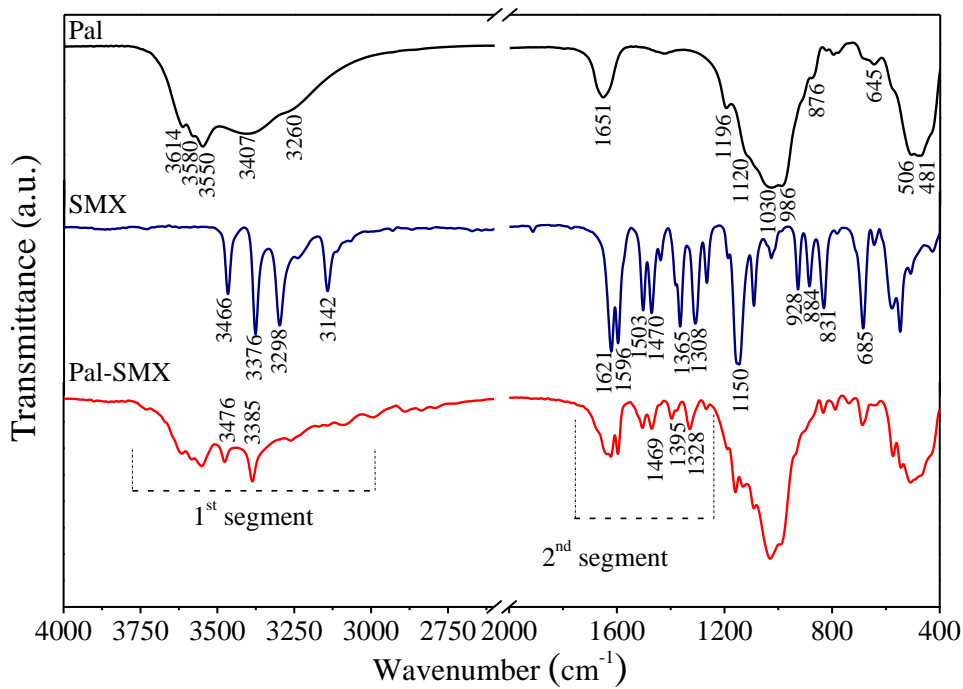

Figure 3. Infrared transmittance spectra for the samples Pal, SMX and the Pal-SMX composite. The characteristics bands and regions are indicated for each one.

The FTIR spectra of SMX exhibited three bands at 3466, 3376 and 3298 cm$^{-1}$, characteristic of the N−H stretching vibrations asymmetric and symmetric of the primary amino group ($v_{NH_2}^{as}$, $v_{NH_2}^{s}$) and the sulfonamide group ($v_{N-H}$), respectively. The band at 3142 cm$^{-1}$ was assigned to the stretching vibrations C−H ($v_{C_{sp^2}-H}$) of the aromatic ring and heterocyclic. The band at 1621 cm$^{-1}$ can be attributed to the combination of several types of vibrations of the SMX molecule: the N−H in-plane bending of the group NH$_2$ ($\delta_{NH_2}$), the stretching vibrations C=N and C=C of the isoxazole ring ($v_{C=N}$), and aromatic ring ($v_{C=C}$), respectively. Associated to the phenyl pattern, two bands at 1596 and 1503 cm$^{-1}$ were identified. The bands at 1470 and 1365 cm$^{-1}$ correspond to the characteristic signals of the isoxazole ring. The stretching frequencies around 1308 and 1150 cm$^{-1}$ were ascribed to asymmetric and symmetric vibrations of the sulfonyl group SO$_2$ ($v_{SO_2}^{as}$ and $v_{SO_2}^{s}$), respectively[20,40,41]. In the low wavenumber region of the spectrum it can be observed at 928 cm$^{-1}$ the overlapping of symmetrical stretching vibration S−N ($v_{S-N}$) with in-plane bending C−H ($\delta_{C-H}$). The bands observed at 884 and 831 cm$^{-1}$ correspond to out-plane bending of C−H ($\gamma_{C-H}$). The signal at 685 cm$^{-1}$ corresponds to the benzene ring deformation[40,41].

In FTIR spectrum of the composite Pal-SMX can be appreciated around 3780 to 3000 cm$^{-1}$ (first segment labeled on the spectrum, see Fig. 3) the overlapping of the bands corresponding to the valence vibration O−Hof Pal together with the valences vibrations N−H and C−H of drug. However, at 3476 and 3385cm$^{-1}$ can be identified the signals assigned to the asymmetric and symmetric stretching vibrations of the group NH$_2$ ($v_{NH_2}^{as}$ and $v_{NH_2}^{s}$). Comparing the SMX spectrum with the composite Pal-SMX, these bands appear displaced at high frequencies, being observed a shift of 10 and 9 cm$^{-1}$, respectively. In the region from 1750 to 1550 cm$^{-1}$ (second segment labeled on the spectrum, see Fig. 3) were observed the signals of the zeolitic and/or adsorbed water present in the Pal and the stretching vibration C=N of the isoxazole ring ($v_{C=N}$), the in-plane bending of the group NH$_2$ ($\delta_{NH_2}$) and the stretching vibrations C=C ($v_{C=C}$) of the aromatic ring of the drug. At 1469 cm$^{-1}$ appears one of the characteristic signals of the isoxazole ring without modification; by contrast the second band is observed at 1395cm$^{-1}$. When comparing this last band in the drug spectrum alone and the composite material, a strong displacement at high frequencies (Δv around 30 cm$^{-1}$) is evidenced. It also occurs for the signal of sulfonyl group. This appears at 1328 cm$^{-1}$ with a displacement of 20 cm$^{-1}$ with respect to the drug spectrum alone. In the area of 1250 to 400 cm$^{-1}$, there is an important overlapping of the characteristic signals of Si–O vibrations of the clay tetrahedrons ($\delta_{Si-nb-Si}$ and $\delta_{O-Si-O}$) with the stretching vibrations (SO$_2$ and S−N), in-plane and out-plane bending C−H ($\delta_{C-H}$ and $\gamma_{C-H}$) of the SMX. These shifts to higher vibration frequencies suggest the existence of interactions between the SMX molecules themselves—via electronegative elements like N of the isoxazole ring and the O of the SO$_2$ group with the hydrogens of the aniline group— and the SMS with the water on the Pal surface. Hence, the most feasible scenario involves the formation of SMX aggregates in its neutral form, i.e., these molecular blocks as a whole interact with the clay surface prevailing the Van de Waals forces and the hydrophobic interactions in the formation of the composite material.

Figure 4, shows the TG/DTG curves for SMX and Pal clay before and after interaction with SMX. TG curve of the raw material (Pal) displays four decomposition steps associated to the loss of the different types of water in the clay structure: adsorbed water, zeolitic water, coordinated water and structural water. The first step begins at 30 ºC and finishes around 197 ºC with a mass loss of 12%. It is the more significant, and it is attributed to the loss of adsorbed water on the surface and zeolitic water. Between 200 and 270 ºC, the mass loss is around of 2%. This second step is ascribed to the residual zeolitic water. From 330º to 540 ºC, the water associated to the coordination sphere of the octahedral cations occupying external positions is gradually lost. In this step the mass loss is approximately of 5%. Within the range 620-710 ºC—accounting for 1% of the initial mass loss—the irreversible structure dihydroxylation (i.e., OH⁻ groups in the octahedral positions)—takes place. At high temperature (> 700 °C) the formation of amorphous Pal occurs [5,29,42,43].

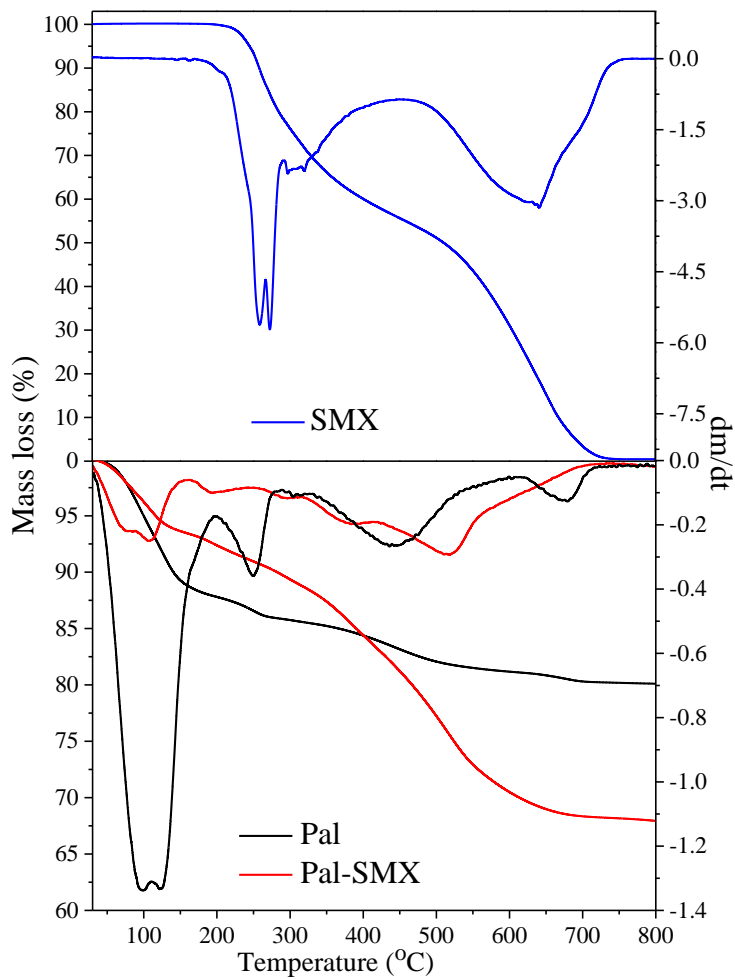

Figure 4. TG/DTG curves for the raw materials (SMX and Pal) and the Pal-SMX composite.

TG/DTG curves of SMX show two decompositions steps (see Fig. 4): the first one begins about 180 °C until 450 °C for a mass loss of 45%, which is attributed to the thermal oxidation of SMX. The second step in the range of 470 to 750°C corresponds to final thermal decomposition of drug (a mass loss about 55%) [20,44]. These results are more clearly illustrated in the DTG curve, where an intense doublet with centers at 259 °C and 273 °C, and other two small signals at 309 °C and 631 °C are visible.

In the DTG curve of the Pal-SMX composite in comparison with the Pal DTG curve changes in the signals can be observed, which occurs at lower temperatures. The first step of mass loss (7%) from 25 ºC to 160 ºC—associated to the adsorbed and zeolitic water— suggests that the presence of SMX (hydrophobic molecule) in the Pal weakens the interactions between the water and clay surface. The second one, associated to a 3% mass loss ofresidual zeolitic water, appears in the interval 160-240 ºC. At higher temperatures, the mass losses corresponding to the oxidation and decomposition of SMX on the sample, as well as to the water removal from of the clay, are evidenced.

*Surface charge*

Surface charge properties of the samples (Pal and Pal-SMX composite) were analyzed from the ZP (see Fig. 5). For Pal, the electrokinetic measurements were negative (−3 to −26 mV) within a broad pH domain and no isoelectric point (point of zero charge, $pH_{pzc}$) was observed. These results indicate that the electrokinetic potential is dominated by the permanent layer charge. Such behavior is consistent with those reported in the literature for 2:1 clays [45,46]. No significant variations were observed in the surface charge for both samples (Pal and Pal-SMX composite), neither qualitatively or quantitatively. Zeta potential values are very similar, indicating the clay colloidal stability after the modification with the drug.

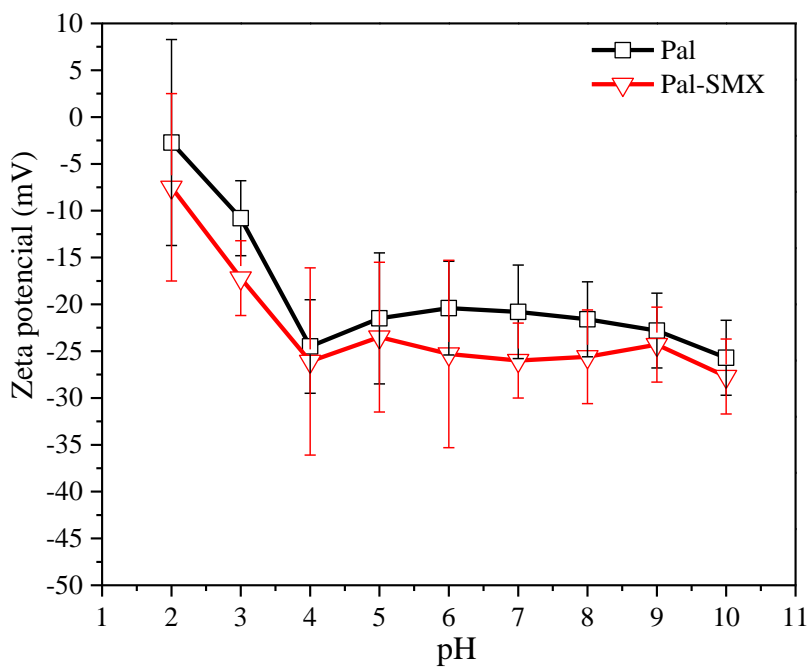

Figure 5. Zeta potential values of Pal and Pal-SMX composite as a function of pH.

*Surface area*

The nitrogen adsorption results showed a reduction of the BET surface area from 115 to 8 m$^2$/g for Pal and SMX-Pal samples, respectively, which indicates a significant decrease of around 93 %. It suggests a SMX adsorption on the clay surface blocking porosity, thus avoiding the N$_2$ diffusion. This sharp reduction is attributed to the existence of SMX molecules in the porous system and it is in agreement with the SMX adsorption results discussed above.

*Morphology analysis*

The morphological structure of the rods of Pal fibers can be observed in the TEM images (see Fig. 6). A large number of Pal rods with a length of a few hundred nanometers, and a diameter of less than 50 nanometers each, can be observed. A rod of palygorskite is formed by several crystals (laths), and various rods form bundles. The laths can appear both aggregated and forming rods, or can remain without aggregation. Generally, the bundle lengths are much longer than the individual laths [47]. As result of the hydrothermal treatment during the preparation of the composite, the Pal structure remained unchanged, as can be seen in Fig. 6b. Comparing the TEM images of the natural Pal with the Pal-SMX composite material, small SMX deposits can be seen on the surface of the Pal fibers after drug interaction (see Fig.6c).

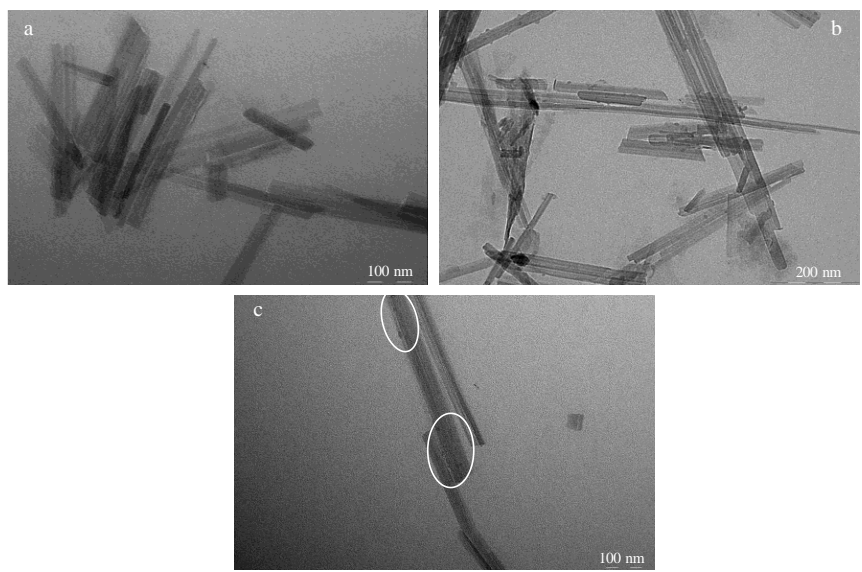

Figure 6. TEM images of Pal (a) and Pal-SMX composite (b and c). Typical regions with some SMX deposits on the surface on the rods are in circles.

*SMX desorption from SMX-Pal composite*

As shown in Fig. 7, the SMX desorption from the composite increases continuously as time goes. The first point corresponds to the drug desorption attached to the clay surface at 24 h, followed by a linear increase. At 96 h, the percentages of SMX desorption is around 43±2%. The results indicate that the SMX adsorption onto the Pal is at least a partially reversible process. The progressive desorption of the drug could be due to the presence of a multilayer of SMX molecules on the Pal surface, where desorption starts by the most external layer. The results suggest the potential reuse of the raw material for a new drug adsorption cycle.

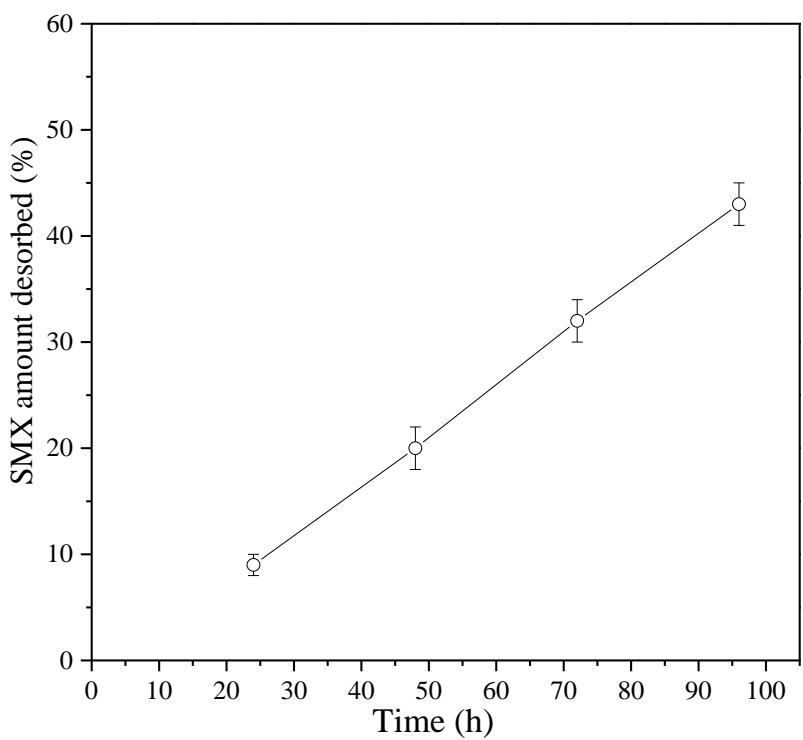

Figure 7: Percent of SMX desorption from Pal-SMX composite in aqueous medium. SMX maxima load onto Pal support was 260 ± 10 mg/g.

**Conclusions**

In this study, the efficient adsorption of the emergent contaminant sulfamethoxazole (SMX) by a Cuban natural palygorskite (Pal) was evaluated. It was demonstrated the adsorption process efficacy at acid pH, room temperature, initial drug concentrationof 3 mg/ml and 30 min of interaction.The structural characterization by XRD allows to confirm the SMX presence on the external surface of the Pal. FTIR analysis indicated the presence of SMX in the clay, as well as the existence of SMX-Pal interactions via Van de Waals forces and hydrophobic interactions in the formation of the composite. In addition, the thermogravimetric analysis suggested that SMX adsorption in the Pal weakens the interactions between the water and clay surface. No significant variations in the zeta potential values were observed for the Pal-SMX sample respect to raw material indicating clay colloidal stability after the modification with the drug. The nitrogen isotherms corroborated the presence of SMS on the Pal surface, evidenced by an important diminish of superficial area of the SMX-Pal regarding Pal. The desorption assays in aqueous medium confirmed the reversibility of the SMX adsorption process onto Pal. The study strongly support the use of Pal as adsorbent of emerging sulfonamide-type contaminants.